\documentclass[runningheads]{llncs}
\usepackage[T1]{fontenc}
\usepackage{graphicx}
\usepackage{stmaryrd}

\usepackage{cmll}
\usepackage{listings}
\usepackage{amsmath}
\usepackage{amssymb}

\usepackage{mathrsfs}
\newcommand{\proves}{\vdash}
\newcommand{\supp}{\Vdash}
\newcommand{\base}[1]{\mathscr{#1}}
\newcommand{\basis}[1]{\mathfrak{#1}}
\newcommand{\setAtoms}{\mathbb{A}}

\newcommand{\argument}[1]{\mathcal{#1}}

\newcommand{\basesup}[1]{\supseteq_{#1}}

\usepackage{xcolor}

\newcommand{\know}{\mathbf{K}}

\newcommand{\baseB}{\base{B}}

\newcommand{\FormToRule}[1]{\llparenthesis {#1} \rrparenthesis}

\usepackage{proof}

\begin{document}
%
\title{On the Logical Content of Knowledge Bases}
%
%
\author{Alexander V. Gheorghiu\inst{1,2}\orcidID{0000-0002-7144-6910} \and \\
Tao Gu\inst{1,2}\orcidID{0000-0001-5749-0758}}
\authorrunning{Gheorghiu \& Gu}
%
\institute{University of Southampton, Southampton SO17 1BJ, UK \and
University College London, London  WC1E 6BT, UK \\
\email{\{a.v.gheorghiu, t.gu\}@soton.ac.uk}
}
\maketitle              
\begin{abstract}
Standard epistemic logics introduce a modal operator $\know$ to represent knowledge, but in doing so presuppose the logical apparatus they aim to explain. By contrast, this paper explores how logic may be derived from the structure of knowledge itself.  We begin from a pre-logical notion of a knowledge base understood as a network of inferential connections between atomic propositions. Logical constants are then defined in terms of what is supported by such a base: intrinsically, by relations that hold within it, and extrinsically, by the behaviour of those relations under extension. This yields a general semantic framework in which familiar systems --- classical, intuitionistic, and various intermediate logics --- arise naturally from different assumptions about the form of knowledge. This offers a reversal of the traditional explanatory order: rather than treating logic as a precondition for the articulation of knowledge, it shows how logical structure can emerge from epistemic organisation.


\end{abstract}
\section{Introduction}
\label{sec:introduction}

The standard treatment of knowledge in logic is modal: one introduces an operator $\know$ such that $\know \phi$ expresses that $\phi$ is known. Systems axiomatizing $\know$ are known as \emph{epistemic logics}~\cite{hintikka1962knowledge}. This framework has proved fruitful, enabling the analysis of common knowledge and multi-agent reasoning, and supplying formal methods for AI, game theory, economics, distributed computing, and natural-language semantics. Philosophical work in this tradition typically concerns the status of principles such as self-reflection ($\know \phi \to \know \know \phi$), rather than the \emph{epistemic content} of formulae themselves. In this sense, such systems are best viewed as accounts of \emph{knowability} rather than of knowledge proper.

By contrast, we argue that knowledge should be taken as conceptually prior to logic. By this we mean that we might begin with a concept of knowledge and determine a logic from it. This is the process typically used in machine intelligence. For example, it is the process used in the traditional approach to AI known as  \emph{logic programming} (LP) --- see  Kowalski~\cite{Kowalski1986,Kowalski1994}. This pragmatic approach to knowledge and reasoning motivates an alternative understanding of `epistemic' logic. 

We begin by introducing a primitive notion of a knowledge base $\base{B}$. This is a structure in which items of information are directly connected. We then define logical signs depending on how one interprets its \emph{logical content}. To illustrate, consider conjunction: What should it mean to say that `$\phi$ and $\psi$’ holds for $\base{B}$? Intuitively, it requires that both $\phi$ and $\psi$ are independently known by $\base{B}$. This case is straightforward, perhaps even trivial. 

More interesting is the case of the conditional: what should it mean to assert that `if $\phi$, then $\psi$'? Intuitively, this says that $\psi$ is supported by $\base{B}$ whenever $\phi$ is. Taken at face value this account collapses into vacuity: the conditional would be satisfied whenever $\phi$ is not supported by $\base{B}$, thereby undermining the hypothetical force of the statement. To properly represent the hypothetical-force of the if-then statement we require something more. We consider extensions $\base{C}$ of $\base{B}$ such that $\psi$ is supported in them whenever $\phi$ is.

To make this analysis precise, in this paper we proceed relative to a simple account of knowledge as an \emph{atomic system}. That is, a collection of atomic propositions structured by some inferential connections. The intuition is developed in several places; see, for example, Piecha and Schroeder-Heister~\cite{Schroeder2016atomic,piecha2017definitional}. To know `All humans are mortal' is to have in mind the inferential connection,
\[
\infer{\text{`$x$ is mortal'}}{\text{`$x$ is a human'}}
\]
as $x$ ranges over the names of all individuals. (Returning to LP, in \textsc{Prolog} the above example is captured by a program containing 
\[
M(x) \mathbin{:\!\!-} H(x)
\] 
where $M$ is the predicate `is mortal', $H$ is the predicate `is human'.) We structure knowledge as this set of inferential connections because some knowledge may be implicit and accessed through internal reasoning. Suppose that an agent Aristotle who knows both the above rule and that `Socrates is a human' (i.e., $H(s)$ where $s$ is `Socrates'). These commitments mean that they also know that `Socrates is mortal'.   

One might object that these inferential connections already presuppose some form of \emph{logic} since they are deductive in character. Importantly though no logical signs are in play here: none have yet been defined. The horizontal bar represents deduction at a meta-theoretic level. Our aim is to \emph{define} logical operators --- implication, conjunction, disjunction, and negation --- by appeal to the \emph{logical content} encoded in these knowledge bases. This is the central work of the paper.

Given a knowledge base $\base{B}$ and an atomic proposition $P$, we write $\proves_{\base{B}} P$ when the commitments contained in $\base{B}$ suffice to derive $P$ --- the details are given in Section~\ref{sec:bes}. An example of an atomic system following the discussion of Aristotle above is as follows:
\[
\infer{M(x)}{H(x)} \qquad \infer{H(s)}{}
\]
as $x$ ranges over the names of individuals. These are natural deduction rules in the sense of Gentzen~\cite{gentzen1969}, but are read \emph{per se} so composed without substitution.

The logical signs are defined by a \emph{support} judgment ($\supp$). In the simplest case, suppose an atomic proposition $P$ follows directly from the commitments of the knowledge base, then $P$ is supported. Thus,
\[
\supp_{\base{B}} P \qquad \mbox{iff} \qquad \proves_{\base{B}} P 
\]
The challenge is to extend this to characterize $\supp_{\base{B}} \phi$ where $\phi$ is a logically-complex formula involving, say, implications, disjunctions, conjunctions, or negations.   

In this formal setup, we may revisit the analysis of implication above to give a formal definition. Recall that intuitively, `if $\phi$, then $\psi$' is supported by a knowledge base $\base{B}$ just in case extension $\base{C} \supseteq \base{B}$ that supports $\phi$ also supports $\psi$. What are these extensions? There are two possibilities. An initial account uses a translation $\FormToRule{-}$ from the logical syntax into inferential form of the knowledge base. Using this, we may represent if-then as the logical sign $\to$ defined as follows:
\[
\supp_{\base{B}} \phi \to \psi \qquad \mbox{iff} \qquad \supp_{\base{B},\FormToRule{\phi}} \psi
\]
The problem with this approach is that such a translation $\FormToRule{\phi}$ might only exist when $\phi$ already has some particular structure (e.g., it is in some clausal form). This restriction makes this clause, perhaps, limited in expressiveness. We may call this the \emph{intrinsic} implication because it speaks of the knowledge itself; for example, the definition of $\phi \to \psi$ considers the situation in which $\phi$ is itself parts of the knowledge base. 

A more cautious alternative is to define a contrasting extrinsic implication $\multimap$. Intuitively, $\phi \multimap \psi$ is supported by $\base{B}$ whenever an arbitrary extension $\base{C} \supseteq \base{B}$ supports $\phi$ it also supports $\psi$:
\[
\supp_{\base{B}} \phi \multimap \psi \qquad \mbox{iff} \qquad \mbox{for any } \base{C} \supseteq \base{B}, \mbox{ if } \supp_{\base{C}} \phi, \mbox{then } \supp_{\base{C}} \psi
\]
Observe that this speaks about the behaviour of the knowledge. That is, $\phi \multimap \psi$ declares the inferential commitment that in any hypothetical scenario where the knowledge is extended in such a way to support $\phi$ one is also committed to $\psi$. Of course, one may expect some parity between intrinsic $\to$ and extrinsic $\multimap$ --- for example, one may expect the latter to be a conservative extension of the former.

In Section~\ref{sec:expressions} we systematically provide intrinsic and extrinsic versions of all the standard connectives --- conjunction, disjunction, and negation.  We then analyse the resulting semantics in Section~\ref{sec:bes}. Perhaps surprisingly, in one setting we recover intuitionistic logic, in another classical logic, and in others intermediate systems that lie between them. This then means that we may read these systems as logics of knowledge; thus, we may say that they are \emph{epistemic} logics in the sense that they express the logical content of knowledge bases. \medskip

This strategy described here of constructing a logic out of justificatory structure (i.e., knowledge represented as a network of inferential links) closely follows Dummett's~\cite{Dummett1991logical} programme. Indeed it places us within the tradition of \emph{proof-theoretic semantics}. This is a long-standing project in modern logic; we defer to Schroeder-Heister~\cite{sep-proof-theoretic-semantics} for a concise summary of its history and development. The central idea is this: the meaning of a logical sign is not fixed by pointing to some external model or mathematical structure; rather, it is fixed  by the inferential role that the sign plays in reasoning. 



\section{Knowledge Bases as Atomic Systems}

We begin with the notion of an \emph{atomic system}. An atomic system records an agent’s commitments concerning the inferential relations between atomic sentences. These are pre-logical in the sense that they contain no logical operators. Formally, atomic systems are sets of atomic rules which are the inferential relations and are represented as natural deduction rules in the sense of Gentzen~\cite{gentzen1969} and generalized by Schroeder-Heister~\cite{schroeder1984natural}.

To illustrate how this works, we consider an example. An agent may or may not know whether or not A(lice) is a parent of B(ob). However, because they understand the relationship between parenthood and relative age they can work that if `A is a parent of B', it must be that `A is older than B'. They can do this because whatever statement $P$ follows from if A is B's father or B's mother also follows A being B's parent. This is represented as an atomic rules as follows:
\[
\infer{P}{\text{A is a parent of B} & \deduce{P}{[\text{A is B's father}]} & \deduce{P}{[\text{A is B's mother}]}}
\]
The $[-]$ marks discharge as usual.
Since the agent also knows
\[
\infer{\text{A is older than B}}{\text{A is B's father}} \qquad \text{ and } \qquad \infer{\text{A is older than B}}{\text{A is B's father}}
\]
they can make the requisite deduction.

Working in a predicate language raises a choice. One may include variables in atomic systems or restrict to closed atoms. Gheorghiu~\cite{Gheorghiu2024FOL} has shown that both formalisms have the same expressive power.
To keep things simple we will \emph{not} include variables.

We assume that we have denumerably many atomic propositions $\setAtoms$.

\begin{definition}[Atomic Rule] \label{def:atomicrule}
 An $n$th-level atomic rule is defined as follows, where $P_1,\dots,P_k,C \in \setAtoms$ and $\Sigma_1,\dots,\Sigma_k$ are (possibly empty) sets of $n$th-level atomic rules:
 \begin{itemize}
  \item[-] A \emph{zeroth-level} atomic rule is a rule
 \[
 \infer{~~C~~}{}
 \]
  \item[-] A \emph{first-level} atomic rule
 \[
 \infer{\,C\,}{\, P_1 & \hdots & P_k \,}
 \]
  \item[-] An \emph{$(n+2)$th-level} atomic rule is a rule 
 \[
 \infer{\,C\,}{ \, \deduce{P_1}{[\Sigma_1]} & \hdots & \deduce{P_k}{[\Sigma_k]} \,}
 \]
 \end{itemize}
\end{definition}

To be clear: premisses may be empty, so that an $m$th-level atomic rule also counts as an $n$th-level atomic rule for any $n>m$.  

\begin{definition}[Atomic System]
    An \emph{atomic system} is a set of atomic rules.
\end{definition}

An atomic system is a set of inferential commitments, but we must be precise about what this amounts to. Heuristically, commitment to the premisses entails commitment to the conclusion; this is the reading of inference adopted by Gentzen~\cite{gentzen1969} --- see also Piecha and Schroeder-Heister~\cite{Schroeder2016atomic,piecha2017definitional}. We formalise this idea by defining a derivability judgement, which specifies when a conclusion is supported by an atomic system together with a set of assumptions. In the general case we proceed as follows:

\begin{definition}[Derivation in an Atomic System]
 Let $\base{A}$ be an atomic system. The set of $\base{A}$-derivations is defined inductively as follows:
 \begin{itemize}
     \item[--] \textsc{Base Case}. If $\base{A}$ contains a zeroth-level rule concluding $C$, then the natural deduction argument consisting only of the node $C$ is an $\base{A}$-derivation.
     Likewise, if $\base{A}$ contains a first-level rule 
     \[
        \infer{\,C\,}{\, P_1 & \hdots & P_k \,}
     \]
     and there is an $\base{A}$-derivation $\mathcal{D}_i$ with hypotheses $\Gamma_i$ and conclusion $P_i$.
    Then the natural deduction argument with root $C$ and immediate sub-trees $\mathcal{D}_1, \dots, \mathcal{D}_k$ is an $\base{A}$-argument from $\Gamma_1 \cup \cdots \cup \Gamma_k$ to $C$.
     \item[--] \textsc{Induction Step}. Suppose $\base{A}$ contains an $(n+2)$th-level rule
    \[
        \infer{\,C\,}{ \, \deduce{P_1}{[\Sigma_1]} & \hdots & \deduce{P_k}{[\Sigma_k]} \,}
    \]
 and suppose that, for each $1 \leq i \leq k$, there is an $\base{A}$-derivation $\argument{D}_i$ with hypotheses $\Gamma_i \cup \Sigma_i$ and conclusion $P_i$.
 Then the natural deduction argument with root $C$ and immediate sub-trees $\argument{D}_1,\dots,\argument{D}_k$ is an $\base{A}$-argument from $\Gamma_1 \cup \dots \cup \Gamma_k$ to $C$.
 \end{itemize}
 An atom $C$ is derivable from $\Gamma$ in $\base{A}$ --- denoted $\Gamma \proves_\base{A} C$ --- iff there is a $\base{A}$-derivation from $\Gamma$ to $C$. 
\end{definition}

Not every form of atomic system needs to be admitted, corresponding to different conceptions of knowledge. To capture this, we introduce the notion of a \emph{basis}, which specifies the class of atomic systems under consideration. Given a fixed basis $\basis{B}$, its elements are called \emph{bases} $\base{B}$. Once a basis is fixed, we always work relative to its bases. Accordingly we introduce the notion of \emph{base-extension}, a restricted version of set-theoretic extension that respects the chosen basis.

\begin{definition}[Basis]
    A \emph{basis} $\basis{B}$ is a set of atomic systems.
\end{definition}

\begin{definition}[Base-extension]
    Given a basis $\basis{B}$, \emph{base-extension} is the smallest relation satisfying
    \[
    \base{Y} \basesup{\basis{B}} \base{X} 
    \quad \mbox{iff} \quad 
    \base{X}, \base{Y} \in \basis{B} \mbox{ and } \base{Y} \supseteq \base{X}.
    \]
\end{definition}

Our semantics for the logical constants will be parameterised by the notion of basis. In particular, we shall consider the following cases, where $\base{A}$ ranges over atomic systems:
\begin{itemize}
\item $\basis{B}_1 := \{\base{A} \mid \mbox{$\base{A}$ is zeroth- or first-level}\}$,  
\item $\basis{B}_2 := \{\base{A} \mid \mbox{$\base{A}$ is zeroth-, first-, or second-level}\}$
\item $\basis{B}_\omega$ comprises all atomic systems.
\end{itemize}

In fact, we will restrict ourselves to the first two only as $\base{B}_2$ and $\base{B}_\omega$ are equivalent in the sense any atomic system in $\basis{B}_\omega$ has an equivalent atomic system in $\base{B}_2$ in terms of derivability. More formally, Stafford et al.~\cite{stafford2025} has shown that if one allows an indefinite supply of fresh atoms, then for each $\base{B} \in \basis{B}_\omega$ one can construct a $\base{B}' \in \basis{B}_{2}$ such that $\proves_{\base{B}} P$ iff $\proves_{\base{B}'} P$ for arbitrary $P$ in the original language (i.e., excluding the fresh atoms). 

We shall henceforth take atomic systems as our model of knowledge bases. A base records the atomic commitments of an agent together with the inferences they acknowledge between them. In this way it represents the informational state of the agent prior to the introduction of logical constants. The subsequent task is to investigate what logical content such a base supports.

\section{Logical Content} \label{sec:expressions}

We have now given a mathematical account of knowledge bases as atomic systems. What remains is to draw out their logical significance. Our aim is to define the logical constants in terms of knowledge itself --- not by appealing to an external semantics of models or valuations, but by seeing how logic arises from the structure and behaviour of knowledge bases. In Section~\ref{sec:introduction}, we saw that there are two distinct attitudes one might take toward what the logical signs express.

One may regard them as describing \emph{intrinsic} relations within the body of knowledge: how the contents of a knowledge base are internally ordered --- this is the view captured by the connective $\to$. This concerns the structure of the current state of knowledge and so the logical signs explain the relations that hold within that state. It thus mirrors the model-theoretic conception based on truth-\emph{in}-a-model after Tarski~\cite{Tarski1936,Tarski2002}. 

Alternatively, one may regard them as expressing \emph{extrinsic} relations: how a body of knowledge behaves when it is extended, revised, or placed in interaction with new information --- this corresponds to the connective $\multimap$. By `behaves' we mean that these connectives are declarations about one's commitments as one's knowledge changes. Taking knowledge to be monotonic (cf. Pollock~\cite{pollock1999contemporary}), the dynamics we have in mind here are \emph{extensions} of the current knowledge base.

Let $\Gamma \supp_{\base{B}} \phi$ denote that $\phi$ is supported when the knowledge $\base{B}$ is extended with inferential commitments that licence one to support the formulae in $\Gamma$. Following the meaning of the extrinsic implication $\multimap$, we may define it as follows: 
\[
\Gamma \supp_{\base{B}} \phi \qquad \mbox{iff} \qquad \text{for any $\base{C} \supseteq \base{B}$, if $\supp_{\base{C}} \psi$ for all $\psi \in \Gamma$, then $\supp_{\base{C}} \phi$}
\]
This definition forms the basis on which we may properly define the extrinsic connectives. 

The \emph{extrinsic} connectives articulate one’s \emph{inferential commitments} under the assumption that their constituent formulas hold. They are distinguished by the way those constituents are structured. The `structures' we have in mind may be organized by the corresponding intrinsic connectives precisely because they are declarations about the state of knowledge. Heuristically, the extrinsic connective~$E$ matches the intrinsic connective~$I$ in the sense that $\phi\,E\,\psi$ holds exactly when, across all possible extensions of the current knowledge base, any atomic proposition~$P$ that would be supported given $\phi\,I\,\psi$ is also supported \emph{in its own right}. This is expressed formally as follows:
\[
\supp_{\base{B}}\phi \, E \, \psi
\quad\text{iff}\quad
\text{for all $\base{C}\supseteq\base{B}$ and all atoms $P$, if }\phi \, I \, \psi \supp_{\base{C}} P,\text{ then }\supp_{\base{C}} P
\]
The extrinsic connectives are therefore characterised not by the \emph{content} of the knowledge base but by the \emph{inferential role} they impose upon it.

\paragraph{Conjunction.} 

A conjunction `$\phi$ and $\psi$' represents that $\phi$ and $\psi$ are both known (relative to a knowledge base $\base{B}$). In Section~\ref{sec:introduction}, we gave the intrinsic reading of this as the familiar form in which $\phi$ and $\psi$ are both independently supported by the knowledge, expressed as follows:
\[
\supp_{\base{B}} \phi \land \psi \qquad \mbox{iff} \qquad \supp_{\base{B}} \phi \text{ and } \supp_{\base{B}} \psi
\]

This leaves open the question of what would be the so-called extrinsic conjunction ($\otimes$). Following the explanation above, $\phi \otimes \psi$ is to be defined by the inferential behaviour  of $\phi \land \psi$. This may be expressed as follows:
\[
\supp_{\base{B}} \phi \otimes \psi \qquad \mbox{iff} \qquad \mbox{for any } \base{C} \supseteq \base{B} \text{ and atom } P, \text{ if } \phi \land \psi \supp_{\base{C}} P \text{ then } \supp_{\base{C}} P
\]
Unpacking $\land$, this can be expressed as follows: 
\[
\supp_{\base{B}} \phi \otimes \psi \qquad \mbox{iff} \qquad \mbox{for any } \base{C} \supseteq \base{B} \text{ and atom } P, \text{ if } \phi, \psi \supp_{\base{C}} P \text{ then } \supp_{\base{C}} P
\]

This formulation captures the idea that $\oplus$ represents a form of conjunction where we acknowledge that the conjunctions hold but without committing to them. In Section~\ref{sec:bes}, we observe that under certain settings these conjunctions collapse into each other.

\paragraph{Disjunction.}

Intuitively, a disjunction `$\phi$ or $\psi$' represents a collection of alternatives. The intrinsic attitude says that one of the disjuncts must hold:
\[
\supp_{\base{B}} \phi \lor \psi \qquad \mbox{iff} \qquad \supp_{\base{B}} \phi \mbox{ or } \supp_{\base{B}} \psi
\]
Following the approach above, the extrinsic disjunction ($\oplus$) is then defined as follows:
\[
\supp_{\base{B}} \phi \oplus \psi \qquad \mbox{iff} \qquad \mbox{for any } \base{C} \supseteq \base{B} \text{ and atom } P, \text{ if } \phi \lor \psi \supp_{\base{C}} P \text{ then } \supp_{\base{C}} P
\]
Unpacking $\lor$, this can be expressed as follows:
\[
\supp_{\base{B}} \phi \oplus \psi \quad \mbox{iff} \quad \mbox{for any } \base{C} \supseteq \base{B} \text{ and atom } P, \text{ if } \phi \supp_{\base{C}} P \text{ and } \psi \supp_{\base{C}} P, \text{ then } \supp_{\base{C}} P.
\]
This formulation captures the idea that $\oplus$ represents a form of disjunction where we acknowledge that one of a set of alternatives holds but without committing to any specific one.

There are many natural examples of cases modelled by such extrinsic disjunctions. Consider, for instance, the following scenario:

There is a lottery with $1000$ tickets. Let the statement that the $i$th ticket wins be $T_i$. Since there are exactly $1000$ tickets, we can assert the disjunction
\[
T_1 \text{ or } T_2 \text{ or } \dots \text{ or } T_{1000}.
\]
The `or' here cannot be the intrinsic disjunction ($\lor$) as we, of course, do not know which ticket will win. Rather it is the extrinsic one; that is, our knowledge is indeed $T_1 \oplus \ldots \oplus T_{1000}$.

What the `or' here means is best understood in terms of behaviour. Suppose that the participants have agreed that, whichever ticket wins, the prize money will be used to buy supplies for an office party. Let $P$ be the statement that the office party will be supplied. Then whoever ticket wins (i.e., whichever $T_i$ holds) it follows that $P$ holds --- that is, $T_i \supp_{\base{B}} P$ for any $i$. The definition of $\oplus$ is exactly such that from this together with the knowledge that at least one ticket wins (i.e., $T_1 \oplus \ldots \oplus T_{1000}$) we know that $P$: the office party will be supplied.


\paragraph{Negation.}

We now turn to negation. To begin, we introduce both an intrinsic and an extrinsic unit for absurdity. Following the pattern established above, the intrinsic one is given by
\[
\supp_{\base{B}} \bot \qquad \text{never holds}
\]
and the extrinsic one by
\[
\supp_{\base{B}} 0 \qquad \text{iff} \qquad \text{for all $P$, } \supp_{\base{B}} P.
\]
These serve as the respective units for negation: a statement is negated by asserting that its assumption leads to absurdity.

Since there are two implications and two units, we obtain four possible negations:
\[
\phi \to \bot, \qquad \phi \to 0, \qquad \phi \multimap \bot, \qquad \phi \multimap 0.
\]
The first of these is vacuous  as it never obtains, but all three of the remainders are meaningful. 

The second (i.e., $\phi \to 0$) expresses that when $\phi$ is added to the knowledge base $\base{B}$, any proposition can be proved. It thus represents an intrinsic notion of negation in the proof-theoretic sense --- that is, as given by \emph{ex falso quodlibet}. 

The third (i.e., $\phi \multimap \bot$) expresses that $\phi$ is not supported by the current knowledge base. It therefore validates the disjunctive syllogism with respect to the intrinsic disjunction:
\[
\text{if } \supp_{\base{B}} \phi \lor \psi \text{ and } \supp_{\base{B}} \phi \multimap \bot, \text{ then } \supp_{\base{B}} \psi.
\]

Finally, the fourth (i.e., $\phi \multimap 0$) gives an \emph{extrinsic} form of disjunctive syllogism. Given $\supp_{\base{B}} \phi \oplus \psi$ and $\supp_{\base{B}} \phi \multimap 0$, we obtain the following conditional:
\[
\text{for any $\base{C} \supseteq \base{B}$ and any atom $P$, if $\psi \supp_{\base{C}} P$, then $\supp_{\base{C}} P$}
\]
This expresses the extrinsic reading of $\phi$ in precisely the sense used to define the extrinsic connectives. It does not ask whether $\phi$ itself is supported, but rather concerns its inferential behaviour: the status of some arbitrary datum (an atom $P$) in a hypothetical situation in which knowing $\phi$ would lead us to support that datum.

\section{Base-extension Semantics} \label{sec:bes}

Assume we have a function $\FormToRule{-}$ from some class of formulae to bases. The logical signs discussed in Section 3 are collected in Figure~\ref{fig:support}, where $\Delta$ is non-empty and for $(\to)$ we have the caveat that $\phi$ belongs to the domain of $\FormToRule{-}$. 

What would be a suitable $\FormToRule{-}$? Given the form of knowledge base, we can choose it as the inverse of the intuitive mapping from bases to sets of formulae. Let $\llbracket - \rrbracket$ be an injection defined as follows:
\[
\begin{cases}
\, \left\llbracket \raisebox{-1mm}{\infer{~~C~~}{}} \right\rrbracket := C \\[2mm]
\left\llbracket \raisebox{-3mm}{\infer{\,C\,}{ P_1 & \hdots & P_n \,} } \right \rrbracket :=  (P_1\land \ldots \land P_n) \to C \\[4mm]
\left\llbracket \raisebox{-5mm}{\infer{\,C\,}{ \, \deduce{P_1}{[\Sigma_1]} & \hdots & \deduce{P_n}{[\Sigma_n]} \,} } \right \rrbracket :=  \big((\llbracket \Sigma_1 \rrbracket \to P_1) \land \ldots \land ( \llbracket \Sigma_n \rrbracket \to P_n) \big) \to C
\end{cases}
\]
 The choice of $\land$ and $\to$ over $\otimes$ and $\multimap$ reflects their intrinsic nature as the formula is put, literally, inside the knowledge base. We set $\FormToRule{-}$ as the left-inverse of this injection. 

\begin{figure}[t]
\hrule 
\vspace{2mm}
 \[
        \begin{array}{l@{\quad}c@{\quad}l@{\quad}r}
            \supp_{\base{B}} P  & \mbox{iff} &   \proves_{\base{B}} P & \mbox{(At)}  \\[1mm]
            \supp_{\base{B}} \phi \to \psi & \mbox{iff} & \supp_{\base{B},\llparenthesis \phi \rrparenthesis} \psi & (\to) \\[1mm]
            \supp_{\base{B}} \phi \multimap \psi & \mbox{iff} & \phi \supp_{\base{B}} \psi & (\multimap) \\[1mm]
            \supp_{\base{B}} \phi \land \psi \hspace{3mm} & \mbox{iff} &    \supp_{\base{B}} \phi \text{ and  } \supp_{\base{B}} \psi& (\land) \\[1mm]
            \supp_{\base{B}} \phi \otimes \psi \hspace{3mm} & \mbox{iff} & \text{for any } \base{C} \supseteq \base{B} \text{ and atom } P, \text{if } \phi, \psi \supp_{\base{C}} P, \text{ then } \supp_{\base{C}} P & (\otimes) \\[1mm]
            \supp_{\base{B}} \bot & & \mbox{never}  & (\bot) \\[1mm]
            \supp_{\base{B}} 0 & \mbox{iff} & \text{for any atom } P, \supp_{\base{B}} P  & (0) \\[1mm]
            \supp_{\base{P}} \phi \lor \psi & \mbox{iff} & \supp_{\base{B}} \phi  \mbox{ or } \supp_{\base{B}} \psi   & (\lor) \\[1mm]
            \supp_{\base{B}} \phi \oplus \psi & \mbox{iff} & \mbox{for any $\base{C} \supseteq \base{B}$ and atom $P$, if $\phi \supp_{\base{Q}} P$ and $\psi \supp_{\base{C}} P$, then $\supp_{\base{C}} P$}  & (\oplus) \\[1mm]
            \hspace{-4mm} \Delta \supp_{\base{B}} \phi & \mbox{iff} & \mbox{$\forall \base{C} \supseteq \base{B}$, if $\supp_{\base{C}} \psi$ for any $\psi \in \Delta$, then $\supp_{\base{C}} \phi$ } &  \mbox{(Inf)} 
        \end{array}
        \]
    \hrule
    \caption{Base-extension Semantics} 
    \label{fig:support}
\end{figure}

We may now ask the following: under this choice of knowledge base as atomic systems, $\FormToRule{-}$ as the left inverse of $\llbracket - \rrbracket$, and these clauses, does this semantics correspond to any known logic? There are several possible answers depending on further details:

The intrinsic fragment of Figure~\ref{fig:support} corresponds to logic programming in the style of Miller~\cite{miller1989logical,Miller91} and Harland~\cite{Harland1991} based on the Harrop formulae. This is a programming language paradigm based on predicate logic: certain clausal formulae (in this case, hereditary Harrop formulae) are read as instructions and computation proceeds via proof-search --- see Kowalski~\cite{Kowalski1974,Kowalski1986}. In this case, the bases $\base{B}$ are \emph{programs} and the judgment $\supp_\base{B} \phi$ says that $\phi$ may be computed from $\base{B}$. This makes sense on the knowledge reading as such programs are sometimes regarded as knowledge bases --- see, for example, Baral and Gelfond~\cite{Baral199473}.

The extrinsic fragment is, perhaps, more surprising. Depending on the basis, one recovers two different but much celebrated logics. Using $\basis{B}_1$ one recovers classical logic but using $\basis{B}_2$ one recovers intuitionistic logic --- see Sandqvist~\cite{Sandqvist2009CL,Sandqvist2015IL} and Gheorghiu~\cite{Gheorghiu2024FOL}. In this setting, the extrinsic conjunction collapses into the intrinsic one:
\[
\supp_{\baseB} \phi \otimes \psi \qquad \mbox{iff} \qquad \supp_{\baseB} \phi \mbox{ and } \supp_{\baseB}  \psi 
\]
--- see Gheorghiu et al.~\cite{GheorghiuGuPym2024IMLL_PTS}. That the change of basis while fixing the semantic clauses should have this effect remains  mysterious.

We may also consider certain hybrid systems that comprise both intrinsic and extrinsic connectives. Stafford et al.~\cite{Stafford2021,stafford2025} have investigated the system comprised of the connectives $\multimap$, $\land$, $\lor$, and $\bot$. If one restricts to $\basis{B}_1$,  they define Stafford's Logic; and, if one takes  $\basis{B}_2$, they determine general inquisitive logic (see Pun\v{c}och\'{a}\v{r}~\cite{Puncochar2016}) --- in short, intuitionistic logic together with the generalized Kreisel-Putnam axiom:
\[
\big(\chi \multimap (\phi \lor \psi)\big) \multimap \big((\chi \multimap \phi) \lor (\chi \multimap \psi)\big)
\]
as $\chi$ ranges over $\lor$-free formulae.

\section{Conclusion} \label{sec:conclusion}

We have presented a framework for analysing the logical content of knowledge bases conceived as atomic systems. Within this setting, the logical constants are defined by the inferential structure of knowledge itself: intrinsically, through relations that hold within a base, and extrinsically, through the behaviour of those relations under extension. The resulting base-extension semantics yields a range of well-known logics --- classical, intuitionistic, and intermediate systems --- depending on the basis adopted. Under this purview, these logics may be understood as epistemic: each expresses a distinct conception of how knowledge is structured and how it persists under revision. Thus, rather than taking logic as prior to knowledge, we may view the traditional logical systems as arising from it as formal articulations of its inferential and dynamic organisation.

This analysis has begun with a deliberately simple conception of a knowledge base as an atomic system. In reality, the structure of knowledge --- whether in human cognition or in artificial agents --- is likely to be far more intricate, or at least could well be conceived as such. Yet even within this minimal setting, we have seen that a seemingly small variation --- the presence or absence of hypotheses in the atomic rules,  the move from $\basis{B}_2$ to $\basis{B}_1$ --- already gives rise to distinct logics, namely the intuitionistic and the classical. It is therefore natural to ask what further logics might emerge when the intrinsic and extrinsic connectives are interpreted over richer, more sophisticated and representative models of knowledge.

\begin{credits}
\subsubsection{\ackname} The authors would like to thank David J. Pym for many thoughtful discussions about proof-theoretic semantics and computational logic that influenced this work, and to Michael Rawson for his input in the formulation of the ideas presented herein.

\subsubsection{\discintname} 
The authors have no competing interests to declare that are relevant to the content of this article.
\end{credits}
%
%
%
\bibliographystyle{splncs04}
\bibliography{bib,gheorghiu}

\end{document}